\documentclass[prl,floatfix,showpacs,preprint,superscriptaddress]{revtex4}
\usepackage{graphicx}
\usepackage{subfigure}
\usepackage{amsmath}
\usepackage{amssymb}
\usepackage{latexsym}
\newcommand{\edd}{\end{document}}
\newcommand{\non}{\nonumber}

\newcommand{\dvr}{\delta v_{r,\,m}}
\newcommand{\dvtheta}{\delta v_{\theta,\,m}}

\begin{document}

\title{ Viscous Fingering-like Instability of 
Cell Fragments }

\author{A.~C. Callan-Jones}
\affiliation{Physicochimie Curie (CNRS-UMR168), Institut Curie, Section de Recherche, 26 rue d'Ulm
75248 Paris Cedex 05 France}
\author{J.-F. Joanny}
\affiliation{Physicochimie Curie (CNRS-UMR168), Institut Curie, Section de Recherche, 26 rue d'Ulm
75248 Paris Cedex 05 France}
\author{J. Prost}
\affiliation{Physicochimie Curie (CNRS-UMR168), Institut Curie, Section de Recherche, 26 rue d'Ulm
75248 Paris Cedex 05 France}
\affiliation{E.S.P.C.I., 10 rue Vauquelin, 75231 Paris Cedex 05, France}
\date{\today}
\begin{abstract}
We present a novel flow instability that can arise in thin films of 
cytoskeletal fluids if the friction with the substrate on which the film lies 
is sufficiently strong.  We consider a two dimensional, membrane-bound 
fragment containing actin filaments that is perturbed from its initially 
circular state, where actin polymerizes at the edge and flows radially inward while
depolymerizing in the fragment.
Performing a linear stability analysis 
of the initial state due to perturbations of the fragment boundary, we 
find, in the limit of very large friction, that the perturbed actin
velocity and pressure fields obey the very same laws governing the 
viscous fingering instability of an interface between immiscible fluids in 
a Hele-Shaw cell. 
A feature of this instability that
is remarkable in the context of cell motility, is that its existence is independent of 
the strength of the interaction between cytoskeletal filaments and myosin
motors, and moreover that it is completely driven by the free energy of actin polymerization at 
the fragment edge.  

\end{abstract}
\pacs{87.17.Jj, 87.17.Rt, 61.30.-v}

\maketitle

Directed motion and shape change allow cells to respond to 
their environment and play  central roles in many biological 
processes such as embryonic development, wound healing, and formation 
of cancer metastases.  Almost universally, crawling of cells on 
a surface or extracellular matrix involves the protrusion of 
a thin leading edge, the lamellipodium, driven by the polymerization 
of the actin cytoskeleton, the adhesion to the substrate 
via specific proteins and molecular motor-enabled contraction 
of the cytoskeleton to translocate the trailing 
cell body~\cite{Pollard_Borisy}.  Remarkably, 
physical units far simpler than eukaryotic cells 
can display self-sustained motion:  the work 
of Ref.~\cite{Verkhovsky_1999} shows that 
nearly flat cell fragments, containing only actin cytoskeleton 
and myosin II motors enclosed by a plasma membrane, are able 
to perform polymerization/contraction-driven motion. Furthermore 
the fragments can spontaneously switch between motile and 
non-motile states.  The observations of Ref.~\cite{Verkhovsky_1999} 
have led us to study theoretically actin-driven motility and 
shape dynamics in these simpler systems with few structural elements 
and few measurable parameters.

The actin cytoskeleton is a highly complex medium:  it is polar 
as actin polymerizes at its ``plus" end, facing the membrane 
abutting the lamellipodium; it is viscoelastic; and it is active 
and driven out of equilibrium by ATP hydrolysis, needed for 
continuous polymerization (treadmilling) and to generate 
myosin motor-induced stresses. 
Recently, a generic hydrodynamic theory has been developed 
to describe active, polar media~\cite{Kruse_generic}. Using a 
small number of phenomenological parameters, it can account for 
a number of motility phenomena due to the coupling of actin filaments 
to myosin activity~\cite{Kruse_asters,Kruse_lamell,Salbreux_oscill}.
The interaction of the cell with its environment is also very 
important in describing motility; for example, it has been 
demonstrated that cells crawling on a heterogeneous substrate tend 
to migrate to regions of greater substrate
adhesion~\cite{Carter_haptotaxis} and and greater 
substrate rigidity~\cite{Lo_durotaxis}.  Cytoskeletal actin 
in a cell or cell fragment is able to transmit forces to 
its substrate through transmembrane proteins
~\cite{Dembo_BiophysJ, Balaban_NCB, Lo_durotaxis}, namely 
integrins, which bind reversibly to the substrate.   
In general, integrins cluster to form focal adhesions 
whose size and mechanical properties are determined by chemical 
and mechanical cues and can regulate the force that they exert.  
If, however, the actin velocity relative to the substrate 
is small compared to $a/\tau$, where $a$ is a molecular size and 
$\tau$ is the average time during which an integrin remains bound, 
then the force exerted by the moving filaments on the substrate 
can be expressed as a friction force, proportional to 
the actin 
velocity~\cite{Tawada_1991,Gerbal_etal, Marcy_etal,Kruse_lamell}.  

\begin{figure}[h]
\centering
\subfigure[] 
{
    \includegraphics[width=3cm]{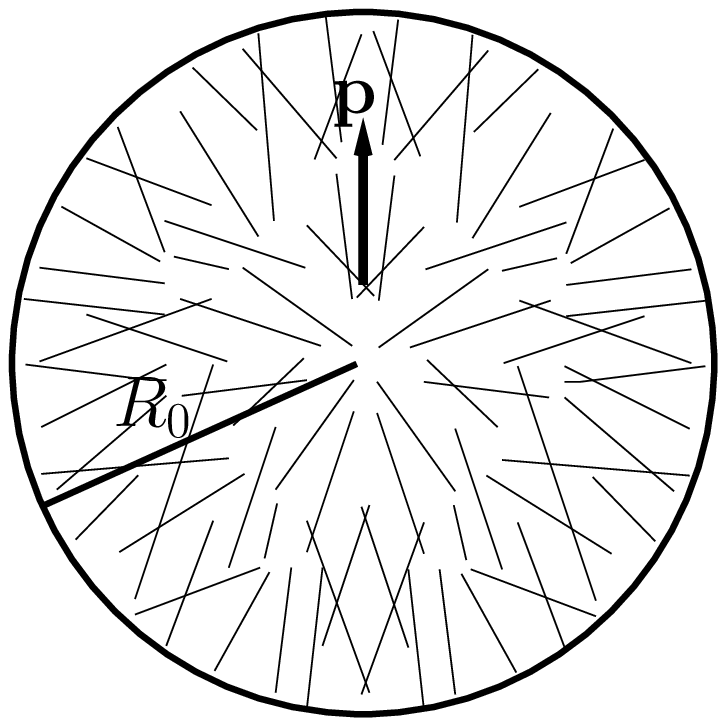}
}
\hspace{3cm}
\subfigure[] 
{
    \includegraphics[width=8cm]{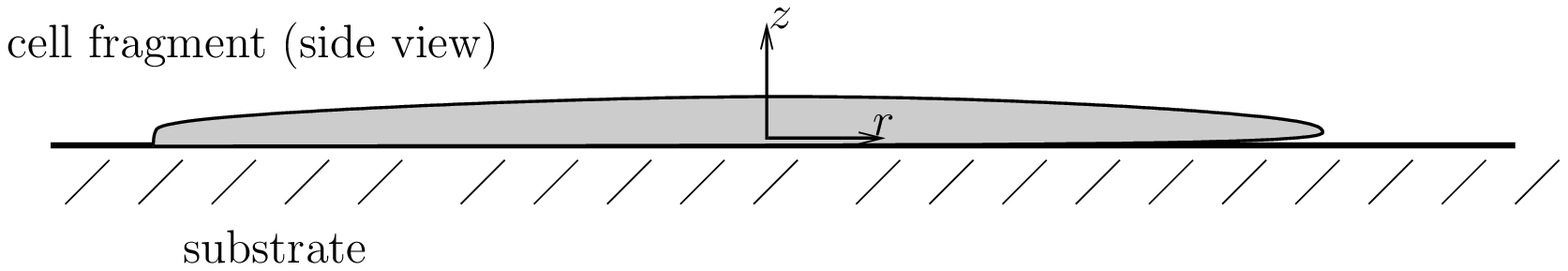}
}
\caption{(a) Schematic drawing of actin cytoskeleton in the unperturbed, radial state.  The direction of average 
actin polarization is radial, and is indicated by $\mathbf{p}=\hat{\mathbf{r}}$.  (b) A side view of the fragment.}
\label{figs:fig1} 
\end{figure}

In this Letter, we demonstrate that polymerization 
and large friction forces are sufficient to destabilize an 
initially stationary, circular cell fragment.  We start by 
considering a very simplified model of actin cytoskeletal 
flow in a cell fragment, as shown schematically in 
Fig.~\ref{figs:fig1}.  The fragment is very thin and there 
is no flow or any spatial dependence in the $z$-direction; 
also we consider that the thickness of the fragment is  constant.  
We further assume that the material in the fragment 
can be modeled as a single fluid: this implies that we 
treat only the flow of actin, and assume that the cytosol 
(including free actin monomers) is stationary relative 
to the substrate.  Furthermore, we treat the cytoskeleton 
as an incompressible liquid, ignoring the elastic 
response that occurs on times shorter than the viscoelastic 
relaxation time.   

Actin polymerization is regulated by  proteins such as those of 
the Wiskott-Aldrich syndrome family (WASP), which localize 
in the cell membrane at the fragment edge~\cite{Alberts}.  
For the purposes of this work, it is sufficient to assume 
that actin polymerizes only at the fragment edge and 
in the direction normal to the boundary.  Newly polymerized 
actin flows away from the fragment edge by treadmilling, 
due the turnover of free actin monomers back to the edge 
for further polymerization that is enabled by actin 
depolymerization in the bulk.    For simplicity, we 
assume that the filament depolymerization is spatially 
uniform and occurs at a rate proportional to the filament density.  

These simplifications imply that the actin flow 
in the unperturbed, circular state induced by localized 
polymerization and uniform depolymerization is imposed by 
the continuity equation $\nabla\cdot (\rho \mathbf{v}_0)=-k_d\rho$, 
where $\rho$ is the actin filament density and $k_d$ is the depolymerization rate.  The assumption 
of incompressibility directly leads to the radially-directed 
treadmilling speed

\begin{equation}
v_0 =-\frac{k_d}{2} r.  
\label{eq:vzero}
\end{equation}

Note that in the stationary state continuity requires 
that $\frac{k_d}{2}R_0=v_p$, where $R_0$ is the unperturbed 
fragment radius and $v_p$ is the polymerization velocity.

The cytoskeleton dynamics is described by the hydrodynamic equations 
for active polar gels of Ref.~\cite{Kruse_generic}, which themselves 
are a generalization of the hydrodynamics of liquid
crystals~\cite{Martin_Pershan_Parodi, DeGennes_Prost} 
modified to account for the coupling between stresses and 
active motors as well as actin polarization and motors.    
We can, however, proceed by considering that the friction 
with the substrate only couples directly to the actin flow 
and not to the polarization.  
The viscous fingering instability that  will be seen shortly 
to be driven by edge polymerization and bulk depolymerization 
in the large friction limit can then be most easily illustrated 
by neglecting the dynamics of the polarization field, 
$\mathbf{p}$, and assuming this quantity is fixed along 
the radial direction so that $\mathbf{p}=\hat{\mathbf{r}}$ throughout.  

Ignoring the dynamics of the polarization field, the constitutive 
laws of Ref.~\cite{Kruse_generic} reduce to those for 
an isotropic, viscous fluid of viscosity $\eta$, augmented 
by an active term in the deviatory stress component $\sigma_{rr}$  
that reflects the myosin-mediated interaction between actin filaments 
that are nearly aligned; that is, $\underline{\underline{\sigma}}=2\eta\underline{\underline{u}}
-\zeta\Delta\mu \hat{\mathbf{r}}\hat{\mathbf{r}}$, where $\eta$ is the actin viscosity; $\underline{\underline{u}}$ is the velocity gradient tensor; $\Delta\mu$ is the chemical potential difference between ATP 
and its hydrolysis products; and where for contractile motors 
the activity coefficient  $\zeta$ is negative~\cite{Kruse_1D_bundle}.  
The constitutive laws are completed at low Reynolds numbers by the force balance
 $\nabla\cdot\underline{\underline{\sigma}}=\nabla P+\xi \mathbf{v}$, where $\xi$ is the friction coefficient between 
the cytoskeletal filaments and the substrate.  Scaling lengths 
by $R_0$, times by $1/k_d$, and stresses and pressures by 
$\eta k_d$ (keeping the same variable names for the new, 
dimensionless quantities), it follows that viscosity and 
friction affect the cytoskeletal dynamics through 
the dimensionless parameter $\lambda^2=\frac{\eta}{\xi R_0^2}$.  
In the limit of very large friction, that is, $\lambda\rightarrow 0$, 
the leading term in the force balance is simply 

\begin{equation}
\mathbf{v}\sim-\lambda^2 \nabla P,
\label{eq:Darcy}
\end{equation}
The velocity satisfies a two dimensional Darcy's law, as it would
for the flow in a Hele-Shaw cell~\cite{Guyon_etal}.  
Based on the available experimental results we find that 
the quantity $\lambda^2$ is in fact quite small.  
Taking a value of $\xi\simeq 10^5$ Pa$\cdot$s/$\mu^2$ ~\cite{Oliver_JCB},
$\eta\simeq 10^4$ Pa$\cdot$s~\cite{Gerbal_etal,Wottawah_2005}, 
and $R_0\simeq 10$ $\mu$m we find $\lambda^2\simeq 10^{-3}$.  

We now perturb the edge of the fragment, so that in terms of 
the polar angle $\theta$ the fragment edge is now at a position
$R(\theta,t)=1+\delta R(\theta,t)$.  For $\delta R(\theta,t)\ll 1$, we perform a linear stability 
analysis by writing
$\delta R(\theta,t)=\sum_{m=1}^{\infty}\delta R_m(t)\cos{(m\theta)}$, 
and similarly for the two components of the perturbed velocity 
field, $\delta v_{r}= \sum_{m=1}^{\infty}\dvr(r,t)\cos{(m\theta)}$ 
and $\delta v_{\theta}= \sum_{m=1}^{\infty}\dvtheta(r,t)\sin{(m\theta)}$,
and the pressure field
$\delta P= \sum_{m=1}^{\infty}\delta P_m(r,t)\cos{(m\theta)}$.  
The $m=0$ mode is excluded from this discussion since it 
is trivially stable because the quantity of actin in the fragment 
is fixed.  In assuming that the depolymerization rate does 
not change as a result of the perturbation and 
that the filament density remains unchanged, it follows that 
$\nabla\cdot \delta \mathbf{v}=0$ and therefore
$\dvtheta=-\frac{1}{m}\frac{d(r\dvr)}{dr}$.  
Applying Eq.~\eqref{eq:Darcy} to the perturbed quantities 
$\delta\mathbf{v}$ and $\delta P$, we find 
that $\nabla^2 \delta P=0$ and therefore 
$\delta P_{m}(r)\sim r^m$ and $\dvr(r,t)=-\dvtheta(r,t)
=A_m(t) r^{m-1}$.  The coefficient $A_m(t)$ can 
be found be imposing the force free condition at the boundary, namely
\begin{equation}
\delta P_m\big|_{r=1}+\delta R_m \frac{dP_0}{dr}\Big|_{r=1}=0,
\end{equation}
leading to $A_m(t)=\frac{m}{2} \delta R_m(t)$.  The growth rate 
of the perturbation modes $\delta R_m(t)$ is obtained by noting 
that, to linear order in $\delta R$, 
\begin{equation}
\frac{d\delta R_m}{dt}\approx \delta v_{r,m}(R_0)+\delta R_m \frac{d v_0}{dr}, 
\label{eq:Rdot}
\end{equation}
which, using the expression for $A_m(t)$, gives $d\delta R_m/dt=\omega_m \delta R_m$,
 where the leading order growth rate, in units of $k_d$, is 
 \begin{equation}
 \omega_m\sim \frac{m-1}{2}+O(\lambda^2).
 \label{eq:omega_leading_order}
 \end{equation}
Note that the mode $m=1$, corresponding to an infinitesimal 
translation of the circular fragment, is marginally stable, 
as required by translational symmetry.  

The linear dispersion relation, $\omega_m$, is a common 
feature to a number of Laplacian growth problems, for example 
the viscous fingering instability that occurs at an interface 
between two immiscible liquids in a Hele--Shaw 
cell~\cite{Bensimon_RMP}.  The physics of the instability
is understood as follows. The pressure gradient at the edge is
$dP_0(r=1)/dr >0$, and a perturbation with $\delta R<0$, 
for example, requires a perturbed pressure $\delta P(r=1)>0$ 
to keep the boundary force-free, to leading order in $\lambda^2$.  
An excess pressure at the edge relative to the fragment center 
($\delta P(r=0)=0$ for $m>0$) drives an inward-directed flow, 
thus amplifying the initial negative perturbation.  

Viscosity, surface tension, and motor activity 
affect the growth rate, $\omega_m$, at
$O(\lambda^2)$.  
In short, one obtains from the radial component of the 
force balance a fourth order ordinary differential 
equation in $r$ for $ u\equiv r\dvr$
\begin{eqnarray}
&&\lambda^2\left[u^{(4)}+\frac{2}{r}u^{(3)}-\frac{(2m^2+1)}{r^2}u''
+\frac{(2m^2+1)}{r^3}u'\right. \non \\[0.3cm]
&&\left.{}+\frac{m^2(m^2-4)}{r^4}u\right]-\left[u''+\frac{1}{r}u'-
\frac{m^2}{r^2}u\right]=0,
\label{eq:forcebalance_u}
\end{eqnarray}
where the primes indicate differentiation with respect to $r$.  
The four boundary conditions required are that 
$\delta \mathbf{v}(r=0)=0$; that the edge of the perturbed 
fragment is force-free, namely 
$\delta \sigma_{nn}(r=1)=0$, where the subscript $n$ refers 
to direction normal to the cell fragment; and that, neglecting 
the viscosity of the fragment's surroundings compared with 
the viscosity of the cytoskeleton, the tangential shear satisfies 
$\delta \sigma_{tn}(r=1)=0$, where $t$ refers to the direction 
tangent to the perturbed fragment.

In the limit $\lambda\rightarrow 0$, 
Eq.~\eqref{eq:forcebalance_u} together with the four boundary 
is a singular perturbation problem.  Following the boundary 
layer techniques of Ref.~\cite{Bender_Orszag}, a uniformly 
convergent approximation to $\dvr$ on the interval $0\le r\le 1$, 
valid to $O(\lambda^2)$,  is 
\begin{equation}
\dvr(r,t)=r^{m-1}\left\{B_0(t)+\lambda^2 \left[B_2 (t)+C_2(t) e^{\frac{(r-1)}{\lambda}}\right]\right\},
\end{equation}
where $B_0(t)=\frac{m}{2}\delta R_m(t)$, is the unperturbed 
growth rate  $A_m(t)$ calculated above which is obtained 
from the boundary condition  $\delta \sigma_{nn}(r=1)=0$ 
at leading order ($O(\lambda^{-2})$); The two other constants 
$B_2(t)$ and $C_2(t)$ are found, respectively, by solving 
$\delta \sigma_{nn}(r=1)=0$ and $\delta\sigma_{tn}(r=1)=0$ 
at next-to-leading order ($O(\lambda^{0})$), giving, 
by way of Eq.~\eqref{eq:Rdot}, a growth rate
\begin{eqnarray}
\hspace{-0.5cm}\omega_{m}&\sim& (m-1)\left\{\frac{1}{2} 
+m \lambda^2\left[\frac{\zeta\Delta\mu}{\eta k_d}
-2 m\right]\right\}+O(\lambda^3).
\label{eq:omega2}
\end{eqnarray}
Equation~\eqref{eq:omega2} shows that the stabilizing effect 
of viscosity is proportional to $m^3$ for large $m$.  
It can be easily seen that the stabilizing effect of the plasma 
membrane tension also scales as $m^3$ (as does the effect of 
interface 
tension in viscous fingering instability in 
the Hele-Shaw cell~\cite{Bensimon_RMP}):  including 
membrane 
tension, the normal stress at the boundary 
satisfies 
a two dimensional Laplace law $\delta \sigma_{nn}(r=1)=-\frac{\gamma}{\eta k_d R_0}
\delta H$, where 
$\gamma$ is the membrane tension and where the $m$th mode 
perturbation in the membrane curvature in the $(r,\theta)$ 
plane (ignoring changes in curvature in the $z$-direction) is 
$\delta H = (m^2-1) \delta R_m \cos{(m\theta)}$.  
The contribution of the membrane tension to $A_2(t)$ is $-\lambda^2
\frac{\gamma}{\eta k_d R_0} m (m^2-1)$.  Taking 
$\gamma \simeq 10^{-4}$ N/m~\cite{Thoumine_cortical_tension} 
as an estimate for the membrane tension and 
$k_d\simeq 1$ s$^{-1}$~\cite{Pollard_1986} it is clear 
that the stabilizing 
effect of membrane tension is negligible compared to that 
of actin viscosity.   

Equation~\eqref{eq:omega2} further shows that the contractile 
effect of the motors is to stabilize the growth of 
perturbations, proportional to $m^2$.  This result 
depends strongly on the assumption, made for simplicity, 
that the filament polarization in the perturbed fragment 
remains everywhere radial and is not a dynamical quantity 
in the problem.  This assumption is questionable, yet it may 
be valid for small $m$, where changes in membrane curvature 
are small and hence any membrane-actin filament coupling 
is unlikely to force a significant reorientation of filaments 
at the leading edge.  In any case, Eq.~\eqref{eq:omega2} 
shows that 
the relative contribution of myosins to $\omega_m$ is proportional 
to $\frac{\zeta\Delta\mu}{\eta k_d}$. Taking 
$\zeta\Delta\mu\simeq -10^3$ Pa~\cite{Kruse_lamell}, this is of 
order $\sim 0.1$, and therefore small compared to the 
viscous contribution at $O(\lambda^2)$.  

Diffusion of free actin monomers also limits the perturbation growth; however, we may
consider for now that the diffusion constant, $D$, is such that $D\gg k_d R_0^2$, so that perturbations
that are  area-preserving at first order in $\delta R$, that is, for $m>1$, do not affect the essentially spatially
uniform monomer density, and hence the polymerization rate, $v_p$.  A more careful accounting 
of the effect of diffusion will be considered in a future publication.

\begin{figure}[h]
\begin{center}
\includegraphics[width=3in,clip=true]{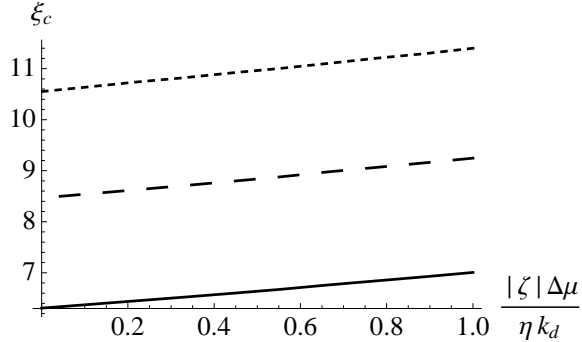}
\caption{Critical value of friction, $\xi_c$, in units of $\eta/ R_0^2$, versus $|\zeta|\Delta\mu/\eta k_d$ for $m=2$ (solid line), $m=3$ (long dashed line), and $m=4$ (short dashed line).  Surface tension, $\sigma/\eta k_d R_0$, is taken to be zero.}
\label{figs:fig2}
\end{center}
\end{figure}

Finally, it might be experimentally useful to have an estimate 
of the critical value of friction, $\xi_c$, for which shape 
perturbations of a cell fragment become unstable.  This critical 
value is defined such that for $\xi<\xi_c$, $\omega_m<0$ 
and for $\xi>\xi_c$, $\omega_m>0$.  It is conceivable that 
one could observe the onset of growing shape perturbations 
by plating cell fragments on surfaces of varying degrees 
of adhesiveness or by culturing fragments from cells 
that have been mutated to weaken or strengthen the binding 
of integrins to the surface~\cite{Burridge_integrins,Tamura_integrins}.
Equation~\eqref{eq:forcebalance_u} can be solved numerically
for different mode numbers $m$ to find the critical value $\xi$ 
where the growth rate becomes positive
as a function of motor strength, $|\zeta|\Delta\mu/\eta k_d$; 
see Fig.~\ref{figs:fig2}.  

The numerical estimates of $\xi_c$ given in Fig.~\ref{figs:fig2} 
are qualitatively consistent with the value obtained by setting 
$\omega_m=0$ in the asymptotic growth rate, Eq.~\eqref{eq:omega2}: 
lower modes are less stable as a function of friction and motor 
activity has a weak effect on the growth of shape perturbations.

In summary, we have found that large substrate friction and 
the pressure field created by treadmilling in an initially circular 
cell fragment render it linearly unstable.  This instability has been 
analyzed here in the limit
$\frac{\eta}{\xi R_0^2}\rightarrow 0$, where it shows 
a close correspondence to the classic viscous fingering instability 
in Hele-Shaw cells. We have also shown by direct calculation 
that the effects of membrane tension, actin viscosity, and 
contractile motors only affect the flow instability 
at next-to-leading order in the friction.  
This instability has the potential to be highly relevant 
to the related biophysical problems of cell shape change 
and cell motility, given that it presents a fundamentally hydrodynamic 
means for cell dynamics, independent of complex biochemical 
signaling and, significantly, of the presence or absence 
of molecular motors.  In future work we would like to study how the instability 
presented here relates to the fragment experiments of Ref.~\cite{Verkhovsky_1999}, in which
a circular, stationary fragment could become anisotropic and motile either spontaneously or due to an external mechanical force; the reverse transition was observed in these experiments as well.  In the context of
the work presented here, the metastability of the fragments of Ref.~\cite{Verkhovsky_1999} might be explained by their friction with the substrate being just below $\xi_c$ for the linear instability of the mode $m=2$.  We are at present considering this possibility, by studying a possible finite amplitude instability that would couple the motile but shape-preserving mode, $m=1$, with the shape changing modes, $m>1$.

\end{document}